\documentclass[draftclsnofoot,onecolumn]{IEEEtran}
\usepackage{fullpage}
\usepackage{citesort}

\usepackage{epsfig}
\usepackage{graphicx}
\usepackage{caption}
\usepackage{subcaption}
\usepackage{fancybox}
\usepackage{color}

\usepackage[cmex10]{amsmath}
\usepackage{amsfonts}
\usepackage{cases}
\usepackage{url}

\usepackage{epstopdf}

\graphicspath{{./} {img/}}
\newtheorem{claim}{Claim}
\newtheorem{corollary}{Corollary}

\begin{document}
%
\title{
\vspace*{-5mm}
Signal Estimation with Additive Error Metrics\\ in Compressed Sensing}
%
%
%

\author{Jin~Tan,~\IEEEmembership{Student Member,~IEEE,}
Danielle~Carmon,
and~Dror~Baron,~\IEEEmembership{Senior Member,~IEEE}
\thanks{This work was supported by the National Science Foundation, grant no. CCF-1217749, and by the U.S. Army Research Office, grant no. W911NF-04-D-0003. Portions of this work were presented at the IEEE Statistical Signal Processing workshop (SSP), Ann Arbor, MI, Aug. 2012~\cite{Tan2012SSP}, and the Information Theory and Applications workshop (ITA), San Diego, CA, Feb. 2013~\cite{TB2013ITA}.}
\thanks{Jin Tan and Dror Baron are with the Department of Electrical and Computer Engineering, North Carolina State University, Raleigh, NC 27695. E-mail: \{jtan; barondror\}@ncsu.edu. Danielle Carmon is with the Department of Cloud Systems Software, IBM, Research Triangle Park, NC 27709. E-mail: dcarmon@us.ibm.com.
}
}
\maketitle \thispagestyle{empty}

\begin{abstract}
Compressed sensing typically deals with the estimation of a system input from its noise-corrupted linear measurements, where the number of measurements is  smaller than the number of input components. The performance of the estimation process is usually quantified by some standard error metric such as squared error or support set error. In this correspondence, we consider a noisy compressed sensing problem with any additive error metric. 
Under the assumption that the relaxed belief propagation method matches Tanaka's fixed point equation,
we propose a general algorithm that estimates the original signal by minimizing the 
additive
error metric defined by the user. The algorithm is a pointwise estimation process, and thus simple and fast. We verify that our algorithm is 
asymptotically optimal,
 and we describe a general method to compute the fundamental information-theoretic performance limit for any additive error metric. We provide 
several example metrics, and give the theoretical performance limits for these cases.
Experimental results show that our algorithm outperforms methods such as relaxed belief propagation (relaxed BP) and compressive sampling matching pursuit (CoSaMP), and reaches the suggested theoretical limits for our example metrics.
\end{abstract}


%
\IEEEpeerreviewmaketitle

\section{Introduction}
\subsection{Motivation}
\label{subsec:motivation}
Compressed sensing~\cite{DonohoCS,CandesUES,CandesRUP} has received a great deal of attention in recent years, because it deals with signal reconstruction problems with far fewer samples than required by the Nyquist rate, greatly reducing the sampling rates required in signal processing applications, such as cameras, medical scanners, and high speed radar~\cite{BaraniukCS2007}.

Consider a linear system,
\begin{eqnarray}
\label{eq:basicSystem}
\mathbf{w=\Phi x},
\end{eqnarray}
where the system input $\mathbf{x}\in\mathbb{R}^N$ is independent and identically distributed (i.i.d.), and the random linear mixing matrix $\mathbf{\Phi}\in\mathbb{R}^{M\times N}$ is sparse and known~(typically $M<N$). The vector $\mathbf{w}\in\mathbb{R}^M$ is called the measurement of $\mathbf{x}$, and is passed through a bank of separable channels characterized by conditional distributions,
\begin{eqnarray}
f_{\mathbf{Y|W}}(\mathbf{y|w})=\prod_{i=1}^M f_{Y|W}(y_i|w_i),
\label{eq:DisChannel}
\end{eqnarray}
where $\mathbf{y}$ is the channel output vector, and $(\cdot)_i$ denotes the $i$th element of a vector.
Note that the channels are general and are not restricted to Gaussian. We observe the channel output $\mathbf{y}$, and want to estimate the original input signal $\mathbf{x}$ from $\mathbf{y}$ and $\mathbf{\Phi}$. 
The remainder of this correspondence follows this system modeled by~\eqref{eq:basicSystem} and~\eqref{eq:DisChannel} (see Rangan~\cite{Rangan2010} for detailed assumptions about the system model).

The performance of the estimation process is often characterized by some error metric that quantifies the distance between the estimated and the original signals. For a signal $\mathbf{x}$ and its estimate $\widehat{\mathbf{x}}$, both of length $N$, the error between them is the summation over the component-wise errors,
\begin{eqnarray}
D(\mathbf{\widehat{x},x})=\sum_{j=1}^N d(\widehat{x}_j, x_j).
\label{eq:distDef}
\end{eqnarray}
For example, if the metric is absolute error, then $d(\widehat{x}_j, x_j)=|\widehat{x}_j-x_j|$; for squared error, $d(\widehat{x}_j, x_j)=(\widehat{x}_j-x_j)^2$.

Squared error is one of the most popular error metrics in various problems, due to many of its mathematical advantages. For example, minimum mean squared error (MMSE) estimation provides both variance and bias information about an estimator~\cite{Grenander1957}, and  in the Gaussian case it is linear and thus often easy to implement~\cite{Levy2008}. However, there are applications where MMSE estimation is inappropriate, for example because it is sensitive to outliers~\cite{Cover91,Webb2002}. Therefore, alternative error metrics, such as mean absolute error (median), mean cubic error, or Hamming distance, are used instead. Considering the significance of various types of error metrics other than squared error, a general estimation algorithm that can minimize any desired error metric is of interest.

\subsection{Related work}
\label{subsec:relatedWork}
As mentioned above, squared error is most commonly used as the error metric in estimation problems given by~\eqref{eq:basicSystem} and~\eqref{eq:DisChannel}. Mean-squared optimal analysis and algorithms were introduced in~\cite{GuoVerdu2005,Guo2006,GuoBaronShamai2009,RFG2012,CSBP2010} to estimate a signal from measurements corrupted by Gaussian noise; in~\cite{GuoWang2007,Rangan2010,RanganGAMP2010}, further discussions were made about the circumstances where the output channel is arbitrary, while, again, the MMSE estimator was put forth. Another line of work, based on a greedy algorithm called {\em orthogonal matching pursuit}, was presented in~\cite{TroppOMP,Cosamp08} where the mean squared error decreases over iterations. Absolute error is also under intense study in signal estimation. For example, an efficient sparse recovery scheme that minimizes the absolute error was provided in~\cite{indyk2008near,berinde2008practical}; in~\cite{CDDNOA}, a fundamental analysis was offered on the minimal number of measurements required while keeping the estimation error within a certain range, and absolute error was one of the metrics concerned. Support recovery error is another metric of great importance, for example because it relates to properties of the measurement matrices~\cite{Wang2010}. 
The authors of~\cite{Tulino2011,Wainwright2009,Wang2010} discussed the support error rate when recovering a sparse signal from its noisy measurements; support-related performance metrics were applied in the derivations of theoretical limits on the sampling rate for signal recovery~\cite{Akcakaya2010,Reeves2011sampling}. 
The readers may notice that previous work only paid attention to limited types of error metrics. What if absolute error, cubic error, or other non-standard metrics are required in a certain application?

\subsection{Contributions}
In this correspondence: ({\em i}) we suggest a pointwise Bayesian estimation algorithm that minimizes an arbitrary additive error metric; ({\em ii}) we prove that the algorithm is optimal; ({\em iii}) we study the fundamental information-theoretic performance limit of an estimation for a given metric; ({\em iv}) we derive the performance limits for 
minimum mean absolute error, minimum mean support error, and minimum mean weighted-support error estimators, and obtain the receiver operating characteristic (ROC) of the modeled system by weighted-support error.
This algorithm is based on the assumption that the relaxed belief propagation (BP) method~\cite{Rangan2010} converges to a set of degraded scalar Gaussian channels~\cite{GuoVerdu2005,GuoBaronShamai2009,RFG2012,GuoWang2007}.
The relaxed BP method is well-known to be optimal for the squared error, while we further show that the relaxed BP method can do more -- because the sufficient statistics are given, other additive error metric can also be minimized with one more simple and fast step.
This is convenient for users who desire to recover the original signal with a non-standard 
additive
error metric. Simulation results show that our algorithm outperforms algorithms such as relaxed BP~\cite{Rangan2010}, which is optimal for squared error, and compressive sampling matching pursuit (CoSaMP)~\cite{Cosamp08}, a greedy reconstruction algorithm. Moreover, we compare our algorithm with the suggested theoretical limits for minimum mean absolute error (MMAE), minimum mean support error (MMSuE), 
and minimum mean weighted-support error (MMWSE),
and illustrate that our algorithm is optimal.

The remainder of the correspondence is arranged as follows: we review relaxed BP in Section \ref{sec:sec2}, and then describe our estimation algorithm and discuss its performance in Section \ref{sec:EstSys}. Simulation results are given in Section \ref{sec:NumSim}, while conclusions appear in Section \ref{sec:Conclusion}. Some mathematical details appear in appendices.

\section{Review of Relaxed Belief Propagation}
\label{sec:sec2}
Before describing the estimation algorithm, a review of the relaxed BP method~\cite{Guo2006,Rangan2010} is helpful. 

\begin{figure}[t]
\centering
\includegraphics[width=50mm]{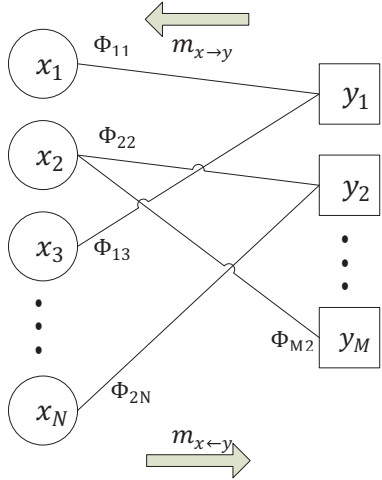}
\vspace*{-5mm}
\caption{
{\small\sl
Tanner graph for relaxed belief propagation.
}
\label{fig:RelaxBP}
}
\vspace*{-5mm}
\end{figure}

Belief Propagation (BP)~\cite{Bishop2006} is an iterative method used to compute the marginals of a Bayesian network. Consider the bipartite graph, called a \textit{Tanner} or \textit{factor} graph, shown in Figure \ref{fig:RelaxBP}, where circles represent random variables (called \textit{variable nodes}), and related variables are connected through functions (represented by squares, called \textit{factor nodes} or \textit{function nodes}) that indicate their dependence. In standard BP, there are two types of messages passed through the nodes: messages from variable nodes to factor nodes, $m_{x\rightarrow y}$, and messages from factor nodes to variable nodes, $m_{y\rightarrow x}$. If we denote the set of function nodes connected to the variable $x$ by $N(x)$, the set of variable nodes connected to the function $y$ by $N(y)$, and the factor function at node $y$ by $\Psi_y$, then the two types of messages are defined as follows~\cite{Bishop2006}:
\begin{eqnarray}
m_{x\rightarrow y}&=&\prod_{k\in N(x)\setminus y}m_{k\rightarrow x},\nonumber\\
m_{y\rightarrow x}&=&\sum_{\ell\in N(y)\setminus x}\Psi_y m_{\ell\rightarrow y}\nonumber.
\end{eqnarray}
Inspired by the basic BP idea described above, the authors of~\cite{Guo2006,GuoWang2007} developed iterative algorithms for estimation problems in linear mixing systems. In the Tanner graph, an input vector $x=[x_1,x_2,...,x_N]$ is associated with the variable nodes (input nodes), and the output vector $y=[y_1,y_2,...,y_M]$ is associated with the function nodes (output nodes). If $\Phi_{ij}\neq0$, then nodes $x_j$ and $y_i$ are connected to an edge $(i,j)$, where the set of such edges $E$ is defined as $E=\{(i,j): \Phi_{ij}\neq0\}$. 

In standard BP methods~\cite{Caire2004,Montanari2006,GuoWang2008}, the distribution functions of $x_j$ and $w_i$ as well as the channel distribution function $f_{Y|W}(y_i|w_i)$ were set to be the messages passed along the graph, but it is difficult to compute those distributions, making the standard BP method computationally expensive. In~\cite{Guo2006}, a simplified algorithm, called \textit{relaxed belief propagation}, was suggested. In this algorithm, means and variances replace the distribution functions themselves and serve as the messages passed through nodes in the Tanner graph, greatly reducing the computational complexity. In~\cite{GuoWang2007,Rangan2010,RanganGAMP2010,Rangan2012}, this method was extended to a more general case where the channel is not necessarily Gaussian.

In Rangan~\cite{Rangan2010}, the relaxed BP algorithm generates two sequences, $q_j(t)$ and $\mu_j(t)$, where $t\in\mathbb{Z}^+$ denotes the iteration number. Under the assumptions that the signal dimension $N\rightarrow\infty$, the iteration number $t\rightarrow\infty$, and the ratio $M/N$ is fixed, the sequences $q_j(t)$ and $\mu_j(t)$ converge to sufficient statistics for the linear mixing channel observation $\mathbf{y}~\eqref{eq:DisChannel}$. More specifically, in the large system limit, the conditional distribution $f(x_j|q_j(t),\mu_j(t))$ converges to the conditional distribution $f(x_j|\mathbf{y})$, where $q_j(t)$ can be regarded as a Gaussian-noise-corrupted version of $x_j$, and $\mu_j(t)$ is the noise variance,
\begin{equation}
\label{eq:1st_scalarGchannel}
q_j(t)=x_j+v_j, 
\end{equation}
where $v_j\sim\mathcal{N}(0,\mu_j(t))$
for $j=1,2,\ldots,N$. It has been shown~\cite{Rangan2010} that $\mu_j(t)$ converges to a fixed point that satisfies Tanaka's equation, which has been analyzed in detail (cf.~\cite{Tanaka2002,GuoVerdu2005,Montanari2006,GuoBaronShamai2009,RFG2012,DMM2009,RanganGAMP2010}). We define the limits of the two sequences,
\begin{eqnarray}
\lim_{t\rightarrow\infty}q_j(t)&=&q_j,\nonumber\\
\lim_{t\rightarrow\infty}\mu_j(t)&=&\mu,\nonumber
\end{eqnarray}
for $j=1,2,\ldots,N$. Note that all the scalar Gaussian channels~\eqref{eq:1st_scalarGchannel} have the same noise variance $\mu$. We now simplify equation~\eqref{eq:1st_scalarGchannel} as follows,
\begin{equation}
q_j=x_j+v_j, 
\label{eq:scalarGchannel}
\end{equation}
where $v_j\sim\mathcal{N}(0,\mu)$
for $j=1,2,\ldots,N$.

\section{Estimation Algorithm}
\label{sec:EstSys}

\subsection{Algorithm}

\begin{figure}[t]
\centering
\includegraphics[width=100mm]{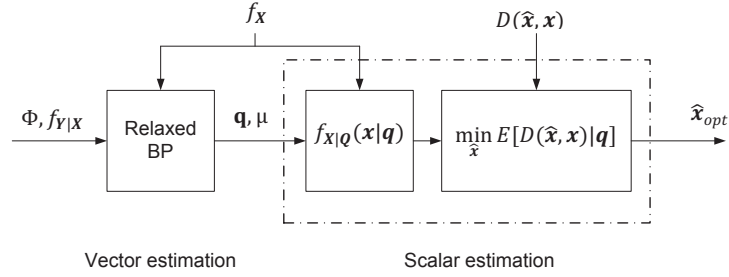}
\vspace*{-5mm}
\caption{
{\small\sl
The structure of the metric-optimal estimation algorithm.
}
\label{fig:EstSys}
}
\vspace*{-5mm}
\end{figure}

The structure of our metric-optimal algorithm is illustrated in the dashed box in Figure~\ref{fig:EstSys}. The inputs of the algorithm are: ({\em i}) a distribution function $f_{\mathbf{X}}(\mathbf{x})$, the prior of the original input $\mathbf{x}$; {(\em ii)} a vector $\mathbf{q}=(q_1,q_2,...,q_N)$, the outputs of the scalar Gaussian channels computed by relaxed BP~\cite{Rangan2010}; ({\em iii}) a scalar $\mu$, the variance of the Gaussian noise in \eqref{eq:scalarGchannel}; and ({\em iv}) an error metric function $D(
\mathbf{\widehat{x},x})$ specified by the user.
The vector $\mathbf{q}$ and the scalar $\mu$ are the outputs of the relaxed BP method by Rangan~\cite{Rangan2010}, and in particular we generate~$\bf{q}$ and~$\mu$ using the software package ``GAMP"~\cite{Rangan:web:GAMP}.

Because the scalar channels have additive Gaussian noise, and that the variances of the noise are all $\mu$, we can compute the conditional probability density function $f_{\mathbf{X|Q}}(\mathbf{x|q})$ from Bayes' rule:
\begin{eqnarray}
f_{\mathbf{X|Q}}(\mathbf{x|q})&=&\frac{f_{\mathbf{Q|X}}(\mathbf{q|x})f_\mathbf{X}(\mathbf{x})}{f_\mathbf{Q}(\mathbf{q})}\nonumber\\
&=&\frac{f_{\mathbf{Q|X}}(\mathbf{q|x})f_\mathbf{X}(\mathbf{x})}{\int f_{\mathbf{Q|X}}(\mathbf{q|x})f_\mathbf{X}(\mathbf{x}) d\mathbf{x}},
\label{eq:BayesRule}
\end{eqnarray}
where
\begin{equation}
f_{\mathbf{Q|X}}(\mathbf{q|x})=\frac{1}{\sqrt{(2\pi\mu)^N}}\exp\left(-\frac{\|\mathbf{q-x}\|_2^2}{2\mu}\right).\nonumber
\end{equation}

Given an error metric $D(\mathbf{\widehat{x},x})$, the optimal estimand $\widehat{\mathbf{x}}_{\text{opt}}$ is generated by minimizing the conditional expectation of the error metric $E[D(\mathbf{\widehat{x},x)|q}]$, which is easy to compute using $f_{\mathbf{X|Q}}(\mathbf{x|q})$: 
\begin{equation}
E[D(\mathbf{\widehat{x},x)|q}]=\int D(\mathbf{\widehat{x},x})f_{\mathbf{X|Q}}(\mathbf{x|q})d\mathbf{x}.\nonumber
\label{eq:conditionalExp}
\end{equation}
Then,
\begin{eqnarray}
\widehat{\mathbf{x}}_{\text{opt}}&=&\arg\min_{\mathbf{\widehat{x}}} E[D(\mathbf{\widehat{x},x)|q}]\nonumber\\
&=&\arg\min_{\mathbf{\widehat{x}}} \int D(\mathbf{\widehat{x},x})f_{\mathbf{X|Q}}(\mathbf{x|q})d\mathbf{x}.
\label{eq:mainAlg}
\end{eqnarray}
The conditional probability $f_{\mathbf{X|Q}}(\mathbf{x|q})$ is separable, because the parallel scalar Gaussian channels~\eqref{eq:scalarGchannel} are separable and $f_{\mathbf{X}}(\mathbf{x})$ is i.i.d. Moreover, the error metric function $D(\mathbf{\widehat{x},x})$~\eqref{eq:distDef} is also separable. Therefore, the problem reduces to scalar estimation~\cite{Levy2008},
\begin{eqnarray}
\widehat{x}_{\text{opt},j}&=&\arg\min_{\widehat{x}_j} E[d(\widehat{x}_j,x_j)|q_j]\nonumber\\
&=&\arg\min_{\widehat{x}_j} \int d(\widehat{x}_j,x_j)f_{x_j|q_j}(x_j|q_j)dx_j,
\label{eq:scalarEst}
\end{eqnarray}
for $j=1,2,\ldots,N$.
Equation~\eqref{eq:scalarEst} minimizes a single-variable function. In Section~\ref{subsec:Eg}, we show how to perform this minimization in three example cases.

\subsection{Theoretical results}
\label{subsec:TheoResult}
Having discussed the algorithm, we now give a theoretical justification for its performance. 
{\em
\begin{claim}
Given the system model described by \eqref{eq:basicSystem}, \eqref{eq:DisChannel} and an error metric $D(\mathbf{\widehat{x},x})$ of the form defined by \eqref{eq:distDef}, as the signal dimension $N\rightarrow\infty$ and the measurement ratio $M/N$ is fixed, the optimal estimand of the input signal is given by
\begin{equation}
\mathbf{\widehat{x}}_{\text{opt}}=\arg\min_{\mathbf{\widehat{x}}} E\left[D(\mathbf{\widehat{x},x)|q}\right],\nonumber
\label{eq:thmEq}
\end{equation}
where the vector entries $\mathbf{q}=(q_1, q_2, \ldots, q_N)$ are the outputs of the scalar Gaussian channels \eqref{eq:scalarGchannel}.
\label{thm:myThm}
\end{claim}
}

The rationale for Claim~\ref{thm:myThm} is as follows. Because the probability density function~$f_{X_j|\mathbf{Y}}(x_j|\mathbf{y})$ is statistically equivalent to~$f_{X_j|Q_j}(x_j|q_j)$ in the large system limit,
once we know the value of $\mu$, estimating each $x_j$ from all channel outputs $\mathbf{y}=(y_1,y_2,...,y_M)$ is equivalent to estimating $x_j$ from the corresponding scalar channel output $q_j$. The relaxed BP method~\cite{Rangan2010} calculates the sufficient statistics $q_j$ and $\mu$. Therefore, an estimator based on minimizing the conditional expectation of the error metric, $E\left(D(\mathbf{\widehat{x},x)|q}\right)$, gives an asymptotically optimal result.

Claim \ref{thm:myThm} states that, in the large system limit, the estimator satisfying~\eqref{eq:mainAlg} is optimal, because it minimizes the conditional expectation of the error metric. The key point in the estimation problem is to obtain the posterior $f_{X_j|\mathbf{Y}}$. Fortunately, the relaxed  BP algorithm provides an asymptotically optimal method to decouple the mixing channels and thus an equivalent posterior $f_{X_j|Q_j}(x_j|q_j)$ can be computed easily, and our algorithm utilizes this convenient feature.

Following Claim \ref{thm:myThm}, we can compute the minimum expected error achievable by {\em any} estimation algorithm for any additive error metric $D(\mathbf{\widehat{x}_{\text{opt}}, x})$. This minimum expected error is the fundamental information-theoretic performance limit of interest in this problem; no estimation algorithm can out-perform this limit. At the same time, we will see in Section~\ref{subsec:Eg} that, for three example error metrics, our BP-based algorithm {\em matches} the performance of the information-theoretic limit, and is thus optimal.

{\em
\begin{claim}
For a system modeled by \eqref{eq:basicSystem}, \eqref{eq:DisChannel}, 
as the signal dimension~$N\to\infty$,
the minimum mean user-defined error (MMUE) is given by
\begin{eqnarray}
\text{MMUE}(f_\mathbf{X},\mu)=\int_{R(\mathbf{Q})}\left(\int_{R(\mathbf{X})}D(\widehat{\mathbf{x}}_{\text{opt}},\mathbf{x})\left(\frac{1}{\sqrt{(2\pi\mu)^{N}}}\exp\left({-\frac{\|\mathbf{q}-\mathbf{x}\|^2}{2\mu}}\right)\right)f_\mathbf{X}(\mathbf{x})d\mathbf{x}\right)d\mathbf{q},\label{eq:MMUE}
\end{eqnarray}
where 
the optimal estimand $\mathbf{\widehat{x}}_{\text{opt}}$ is determined by \eqref{eq:mainAlg},
$R(\cdot)$ represents the range of a variable, and $\mu$ is the variance of the noise of the scalar Gaussian channel \eqref{eq:scalarGchannel}.
\label{thm:BoundExpression}
\end{claim}
}

Equation~\eqref{eq:MMUE} can be derived in the following steps.
\begin{eqnarray}
\allowdisplaybreaks
&&\text{MMUE}(f_\mathbf{X},\mu)=E[D(\widehat{\mathbf{x}}_{\text{opt}},\mathbf{x})]\nonumber\\
&=&\int_{R(\mathbf{Q})}E_\mathbf{Q}\bigg[E[D(\widehat{\mathbf{x}}_{\text{opt}},\mathbf{x})|\mathbf{q}]\bigg]f_{\mathbf{Q}}(\mathbf{q})d\mathbf{q}\nonumber\\
&=&\int_{R(\mathbf{Q})}E[D(\widehat{\mathbf{x}}_{\text{opt}},\mathbf{x})|\mathbf{q}]f_\mathbf{Q}(\mathbf{q})d\mathbf{q}\nonumber\\
&=&\int_{R(\mathbf{Q})}\left(\int_{R(\mathbf{X})}D(\widehat{\mathbf{x}}_{\text{opt}},\mathbf{x})f_{\mathbf{X|Q}}(\mathbf{x|q})d\mathbf{x}\right)f_\mathbf{Q}(\mathbf{q})d\mathbf{q}\nonumber\\
&=&\int_{R(\mathbf{Q})}\left(\int_{R(\mathbf{X})}D(\widehat{\mathbf{x}}_{\text{opt}},\mathbf{x})\frac{f_{\mathbf{Q|X}}(\mathbf{q|x})f_\mathbf{X}(\mathbf{x})}{f_\mathbf{Q}(\mathbf{q})}d\mathbf{x}\right)f_\mathbf{Q}(\mathbf{q})d\mathbf{q}\nonumber\\
&=&\int_{R(\mathbf{Q})}\left(\int_{R(\mathbf{X})}D(\widehat{\mathbf{x}}_{\text{opt}},\mathbf{x})f_{\mathbf{Q|X}}(\mathbf{q|x})f_\mathbf{X}(\mathbf{x})d\mathbf{x}\right)d\mathbf{q}\nonumber\\
&=&\int_{R(\mathbf{Q})}\left(\int_{R(\mathbf{X})}D(\widehat{\mathbf{x}}_{\text{opt}},\mathbf{x})\frac{1}{\sqrt{(2\pi\mu)^{N}}}\exp\left({-\frac{\|\mathbf{q}-\mathbf{x}\|^2}{2\mu}}\right)f_\mathbf{X}(\mathbf{x})d\mathbf{x}\right)d\mathbf{q}\nonumber.
\end{eqnarray}

Using both claims, we further analyze the estimation performance limits for three example error metrics in Sections~\ref{subsec:Eg}.

\subsection{Examples}
\label{subsec:Eg}

\subsubsection{Absolute error}
\label{subsub:Abs}

Because the MMSE is the mean of the conditional distribution, the outliers in the set of data may corrupt the estimation, and in this case the minimum mean absolute error (MMAE) is a good alternative. For absolute error, $d_{\text{AE}}(\widehat{x_j},x_j)=|\widehat{x_j}-x_j|$, and we have the following corollary describing the performance limit of an MMAE estimator, where the proof is given in Appendix \ref{appen:MMAEProof}.

{\em
\begin{corollary}
\label{coro:MMAE}
For a system modeled by \eqref{eq:basicSystem}, \eqref{eq:DisChannel}, 
as the signal dimension~$N\to\infty$,
the minimum mean absolute error (MMAE) estimator achieves
\begin{equation}
\text{MMAE}(f_{\mathbf{X}},\mu)=N\int_{-\infty}^{+\infty}\left(\int_{-\infty}^{\widehat{x}_{j,\text{MMAE}}} (-x_j)f_{X_j|Q_j}(x_j|q_j)dx_j+\int_{\widehat{x}_{j,\text{MMAE}}}^{+\infty} x_j f_{X_j|Q_j}(x_j|q_j)dx_j\right)f_{Q_j}(q_j)dq_j,
\label{eq:MMAEBound}
\end{equation}
where $x_j$ (respectively, $q_j$) is the input (respectively, output) of the scalar Gaussian channel \eqref{eq:scalarGchannel}, $\widehat{x}_{j,\text{MMAE}}$ satisfies $\int_{\widehat{x}_{j,\text{MMAE}}}^{+\infty}f_{X_j|Q_j}(x_j|q_j)dx_j=\frac{1}{2}$, and $f_{X_j|Q_j}(x_j|q_j)$ is a function of $f_{X_j}$ following \eqref{eq:BayesRule}.
\end{corollary}
}

\subsubsection{Support recovery error}
\label{subsub:support}

In some applications in compressed sensing, correctly estimating the locations where the data has non-zero values is almost as important as estimating the exact values of the data; it is a standard model selection error criterion~\cite{Wang2010}. The process of estimating the non-zero locations is called \textit{support recovery}. Support recovery error is defined as follows, and this metric function is discrete,
\begin{eqnarray}
d_{\text{support}}(\widehat{x}_j,x_j)=\text{xor} (\widehat{x_j}, x_j),\nonumber
\end{eqnarray}
where
\begin{equation}
\text{xor}(\widehat{x}_j, x_j)=
\begin{cases}
0, \text{ if }x_j=0 \text{ and }\widehat{x}_j=0\\
0, \text{ if }x_j\neq0 \text{ and }\widehat{x}_j\neq0\\
1, \text{ if }x_j=0 \text{ and }\widehat{x}_j\neq0\\
1, \text{ if }x_j\neq0 \text{ and }\widehat{x}_j=0
\end{cases}\nonumber.
\end{equation}

{\em
\begin{corollary}
\label{coro:MMSuE}
For a system modeled by \eqref{eq:basicSystem}, \eqref{eq:DisChannel}, where $f_\mathbf{X}$ is an i.i.d. sparse Gaussian prior such that~$\Pr(X_j\neq0)=p$ and~$X_j\neq0\sim\mathcal{N}(0,\sigma^2)$, 
as the signal dimension~$N\to\infty$,
the minimum mean support error (MMSuE) estimator achieves
\begin{equation}
\text{MMSuE}(f_{\mathbf{X}},\mu)=N\cdot(1-p)\cdot\text{erfc}\left(\sqrt{\frac{\tau}{2\mu}}\right)+N\cdot p\cdot\text{erf}\left(\sqrt{\frac{\tau}{2(\sigma^2+\mu)}}\right),
\label{eq:supportBound}
\end{equation}
where
\begin{equation}
\tau=2\cdot\frac{\sigma^2+\mu}{\sigma^2/\mu}\cdot\ln\left(\frac{(1-p)\sqrt{\sigma^2/\mu+1}}{p}\right).\nonumber
\end{equation}
\end{corollary}
}

Corollary \ref{coro:MMSuE} is proved in Appendix~\ref{appen:MMSuEProof}.

\subsubsection{Weighted-support error}
\label{subsub:ROC}
In Section~\ref{subsub:support}, we put equal weights on the error patterns {\em (i)} $\widehat{x}_j\neq0$ while $x_j=0$; and {\em (ii)} $\widehat{x}_j=0$ while $x_j\neq0$. In this section, we further put unequal weights on these two error patterns, and the {\em receiver operating characteristic} (ROC) curve~\cite{Levy2008} is obtained. We first define {\em false positive} error $d_{\text{FP}}$ as
\begin{equation}
d_{\text{FP}}(\widehat{x}_j,x_j)=d(\widehat{x}_j=1,x_j=0)=1,\nonumber
\label{eq:FP}
\end{equation}
and {\em false negative} error $d_{\text{FN}}$ as
\begin{equation}
d_{\text{FN}}(\widehat{x}_j,x_j)=d(\widehat{x}_j=0,x_j=1)=1.\nonumber
\label{eq:FN}
\end{equation}
Else the patterns coincide,
\begin{equation}
d(\widehat{x}_j=0,x_j=0)=d(\widehat{x}_j=1,x_j=1)=0.\nonumber
\end{equation}
We then define {\em weighted-support error} as
\begin{equation}
d_{\text{w\_support}}(\widehat{x}_j,x_j)=\beta\cdot d_{\text{FP}}(\widehat{x}_j,x_j)+(1-\beta)\cdot d_{\text{FN}}(\widehat{x}_j,x_j),
\label{eq:W_support}
\end{equation}
where $0\le\beta\le 1$, and we set $d_{\text{w\_support}}(\widehat{x}_j,x_j)$ as the error metric that we want to minimize. The false positive rate (or {\em false alarm rate}) is defined as $\Pr(\widehat{x}_j\neq0|x_j=0)$, and the false negative rate (or {\em misdetection rate}) is defined as $\Pr(\widehat{x}_j=0|x_j\neq0)$.

{\em
\begin{corollary}
\label{coro:MMWSE}
For a system modeled by \eqref{eq:basicSystem}, \eqref{eq:DisChannel}, where $f_\mathbf{X}$ is an i.i.d. sparse Gaussian prior such that~$\Pr(X_j\neq0)=p$ and~$X_j\neq0\sim\mathcal{N}(0,\sigma^2)$, 
as the signal dimension~$N\to\infty$,
\begin{enumerate}
\item
 the minimum mean weighted-support error (MMWSE) estimator achieves
\begin{equation}
\text{MMWSE}(f_{\mathbf{X}},\mu)=N\beta(1-p)\cdot\text{erfc}\left(\sqrt{\frac{\tau'}{2\mu}}\right)+N(1-\beta) p\cdot\text{erf}\left(\sqrt{\frac{\tau'}{2(\sigma^2+\mu)}}\right),
\label{eq:w_supportBound}
\end{equation}
where
\begin{equation}
\tau'=2\cdot\frac{\sigma^2+\mu}{\sigma^2/\mu}\cdot\ln\left(\frac{\beta(1-p)\sqrt{\sigma^2/\mu+1}}{(1-\beta)p}\right).\nonumber
\end{equation}
\item
The false positive rate is
\begin{equation}
\Pr(\widehat{x}_j\neq0|x_j=0)=\text{erfc}\left(\sqrt{\frac{\tau'}{2\mu}}\right),
\label{eq:FAR}
\end{equation}
and the false negative rate is
\begin{equation}
\Pr(\widehat{x}_j=0|x_j\neq0)=\text{erf}\left(\sqrt{\frac{\tau'}{2(\sigma^2+\mu)}}\right).
\label{eq:MDR}
\end{equation}
\end{enumerate}
\end{corollary}
}

The proof of Corollary~\ref{coro:MMWSE} is provided in Appendix~\ref{appen:W_SuEProof}.

It is shown in Corollary~\ref{coro:MMWSE} that $\Pr(\widehat{x}_j\neq0|x_j=0)$ and $\Pr(\widehat{x}_j=0|x_j\neq0)$ vary when the value of $\beta$ varies. 
Moreover, Comparing equation~\eqref{eq:w_supportBound} to~\eqref{eq:supportBound} in Corollary~\ref{coro:MMSuE}, the only difference is that $\tau$ is replaced by $\tau'$, which is the {\em decision threshold} that determines whether an estimand is zero or nonzero.
That said, putting different weights on false positive error and false negative error is analogous to tuning the decision threshold, and thus trading off between the false alarm rate and the misdetection rate~\cite{Levy2008}. An ROC curve is shown in Section~\ref{sec:NumSim}.

\section{Numerical Results}
\label{sec:NumSim}

Some numerical results are shown in this section to illustrate the performance of our estimation algorithm when minimizing a user-defined error metric. The Matlab implementation of our algorithm can be found at \url{http://people.engr.ncsu.edu/dzbaron/software/arb_metric/}.

We test our estimation algorithm on two linear systems modeled by \eqref{eq:basicSystem} and \eqref{eq:DisChannel}: ({\em i}) Gaussian input and Gaussian channel; ({\em ii}) Weibull input and Poisson channel. In both cases, the input's length $N$ is 10,000, and its sparsity rate is $3\%$, meaning that the entries of the input vector are non-zero with probability $3\%$, and zero otherwise. The matrix $\mathbf{\Phi}$ we use is Bernoulli($0.5$) distributed, and is normalized to have unit-norm rows. In the first case, the non-zero input entries are $\mathcal{N}(0,1)$ distributed, and the Gaussian noise is $\mathcal{N}(0,3\cdot 10^{-4})$ distributed, i.e., the signal to noise ratio (SNR) is 20 dB. In the second case, the non-zero input entries are Weibull distributed,
\begin{eqnarray}
f(x_j;\lambda,k)=
\begin{cases}
\frac{k}{\lambda}\left(\frac{x_j}{\lambda}\right)^{k-1}e^{-(x_j/\lambda)^k}&x_j\ge 0\\
0&x_j<0
\end{cases},\nonumber
\end{eqnarray}
where $\lambda=1$ and $k=0.5$. The Poisson channel is
\begin{eqnarray}
f_{Y|W}(y_i|w_i)=\frac{(\alpha w_i)^{y_i}e^{-(\alpha w_i)}}{y_i!},\quad\text{ for all }i\in\{1,2,\ldots,M\},\nonumber
\end{eqnarray}
where the scaling factor of the input is $\alpha=100$.

In order to illustrate that our estimation algorithm is suitable for reasonable error metrics, we considered absolute error and two other non-standard metrics: 
\begin{eqnarray}
\label{eq:OtherMetric}
\text{Error}_p=\sum_{j=1}^N |\widehat{x}_j-x_j|^p,\nonumber
\end{eqnarray}
where $p=0.5$ or $1.5$.

\begin{figure}[t]
\centering
	\begin{subfigure}[t]{0.3285\textwidth}
		\centering
		\includegraphics[width=56mm]{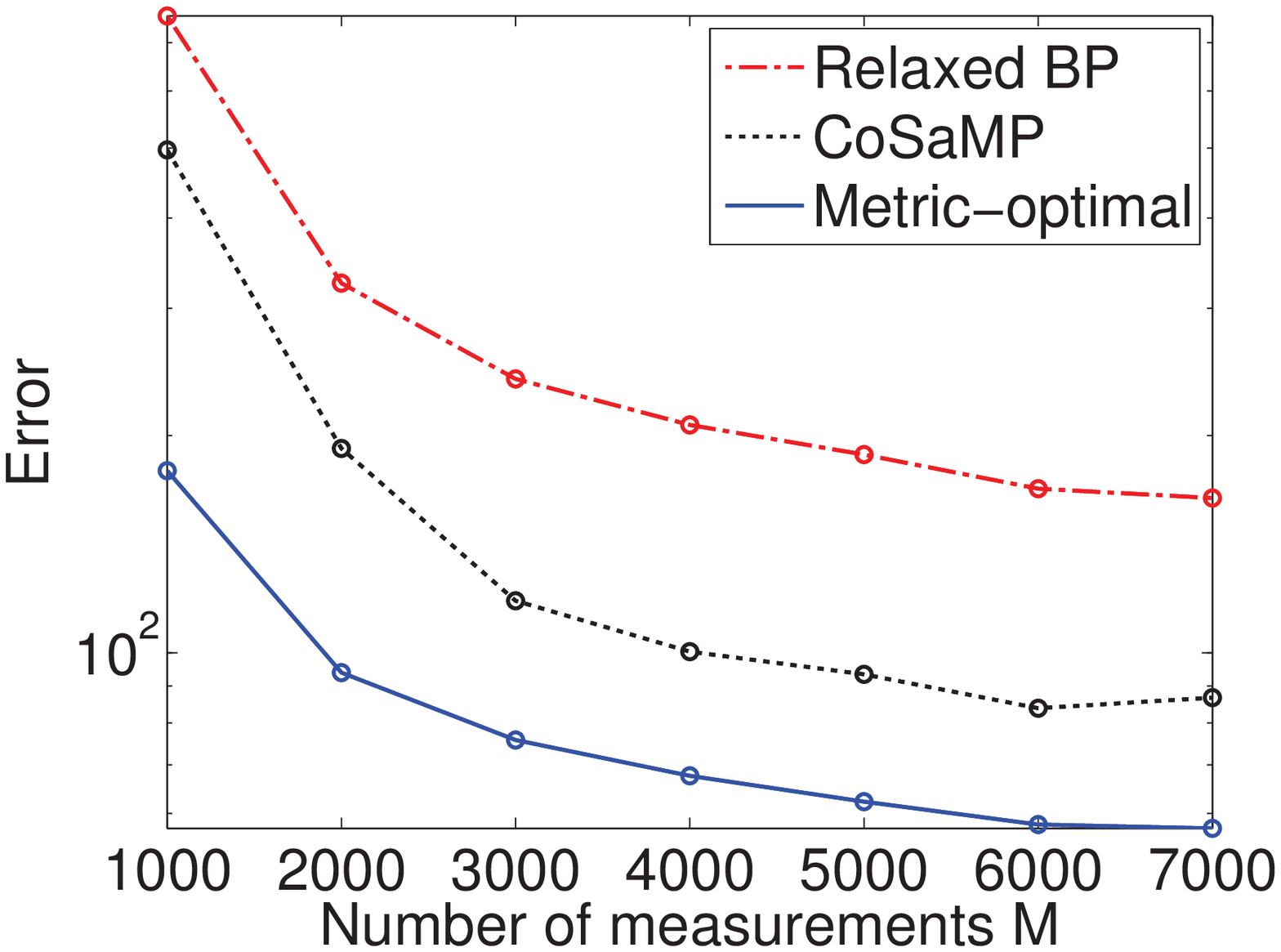}
\caption{\footnotesize\sl$D(\mathbf{\widehat{x},x})=\sum_{j=1}^N|\widehat{x}_j-x_j|^{0.5}$.}
		\label{fig:plotG_1}
	\end{subfigure}
	\begin{subfigure}[t]{0.3285\textwidth}
		\centering
		\includegraphics[width=56mm]{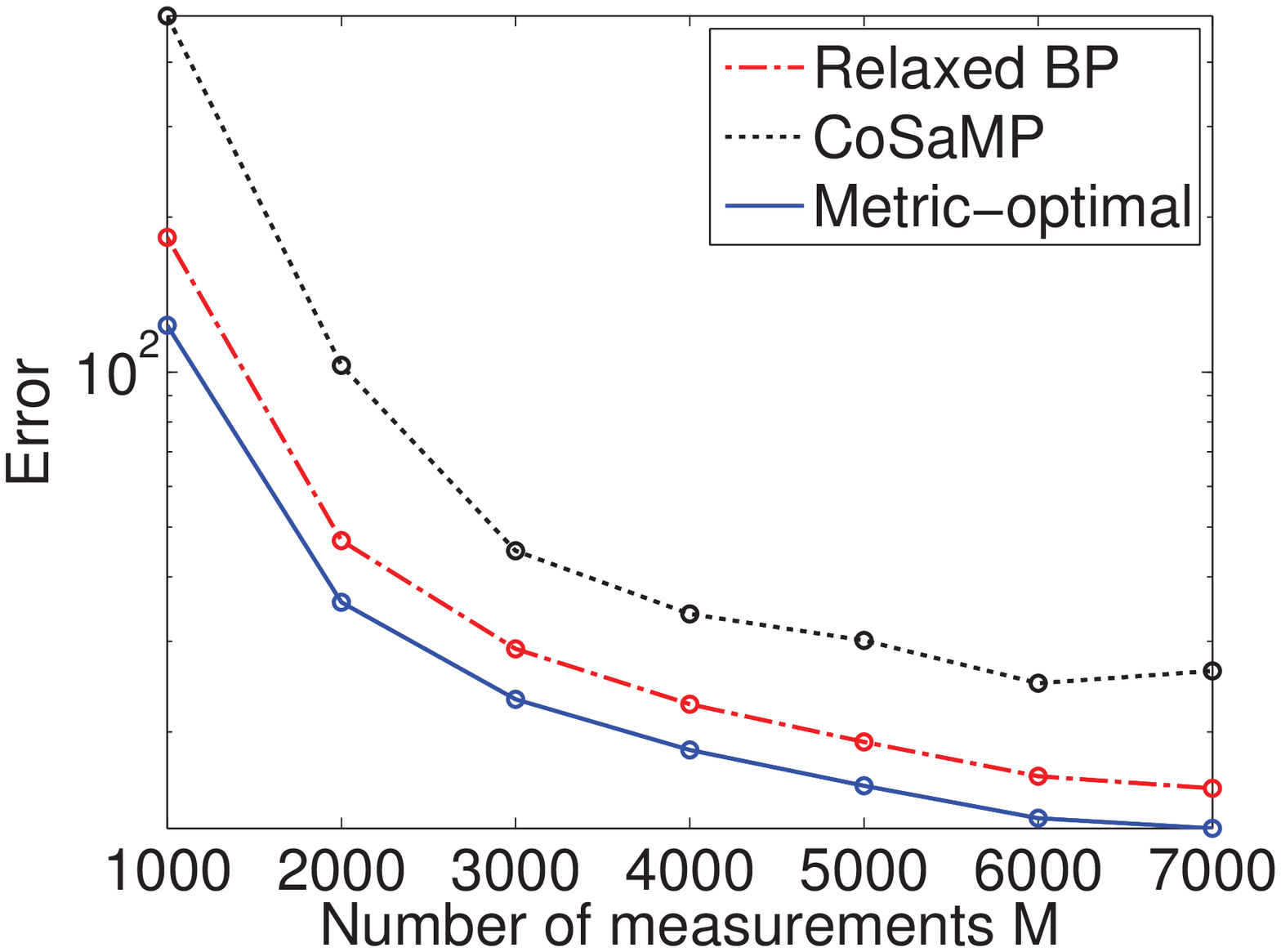}
\caption{\footnotesize\sl$D(\mathbf{\widehat{x},x})=\sum_{j=1}^N|\widehat{x}_j-x_j|$.}
		\label{fig:plotG_2}
	\end{subfigure}
	\begin{subfigure}[t]{0.3285\textwidth}
		\centering
		\includegraphics[width=56mm]{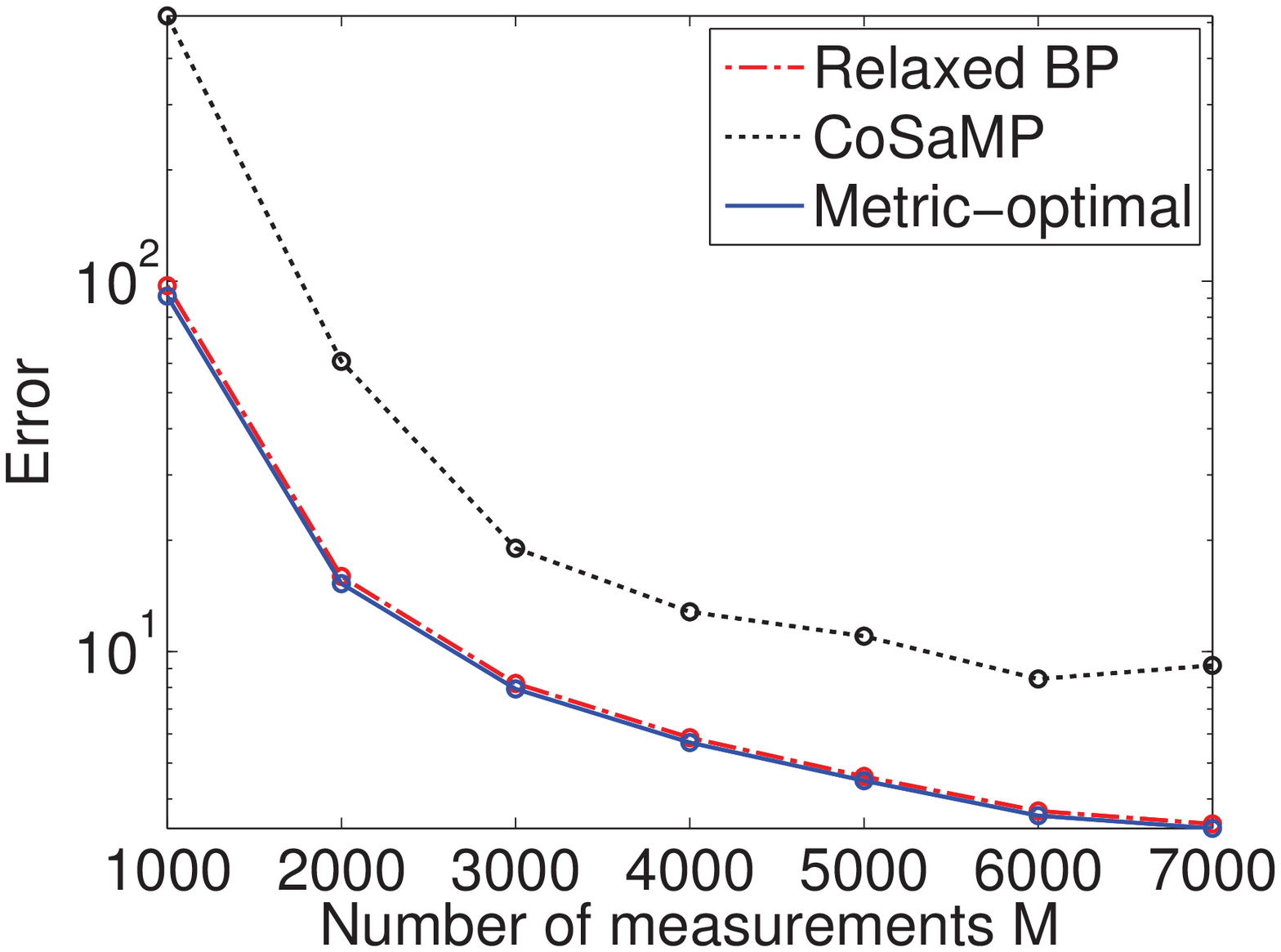}
\caption{\footnotesize\sl$D(\mathbf{\widehat{x},x})=\sum_{j=1}^N|\widehat{x}_j-x_j|^{1.5}$.}
		\label{fig:plotG_3}
	\end{subfigure}
\caption{
{\small\sl
Comparison of the metric-optimal estimation algorithm, relaxed BP, and CoSaMP. (Sparse Gaussian input and Gaussian channel; sparsity rate $=3\%$; input length $N=10,000$; SNR $=20$ dB.)
}
\label{fig:PlotG}
}
\vspace*{-5mm}
\end{figure}

\begin{figure}[t]
\centering
\includegraphics[width=70mm]{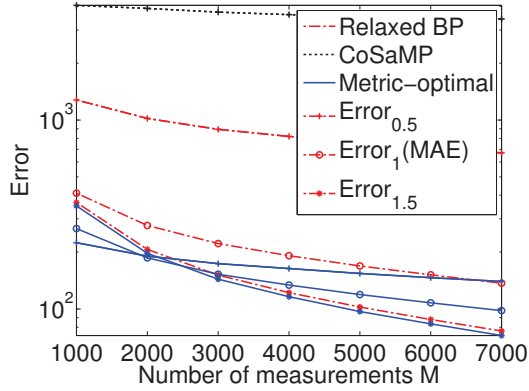}
\vspace*{-5mm}
\caption{
{\small\sl
Comparison of the metric-optimal estimation algorithm, relaxed BP, and CoSaMP. The ``MAE" and the ``$\text{Error}_{1.5}$" lines for ``CoSaMP" appear beyond the scope of vertical axis. (Sparse Weibull input and Poisson channel; sparsity rate $=3\%$; input length $N=10,000$; input scaling factor $\alpha=100$.)
}
\label{fig:PlotW}
}
\vspace*{-5mm}
\end{figure}

We compare our algorithm with the relaxed BP~\cite{Rangan2010} and CoSaMP~\cite{Cosamp08} algorithms. In Figure \ref{fig:PlotG} and Figure \ref{fig:PlotW}, lines marked with ``metric-optimal" present the errors of our estimation algorithm, and lines marked with ``Relaxed BP" (respectively, ``CoSaMP") show the errors of the relaxed BP (respectively, CoSaMP) algorithm. Each point in the figure is an average of 100 experiments with the same parameters. Because the Poisson channel is not an additive noise channel and is not suitable for CoSaMP, the ``MAE" and the ``$\text{Error}_{1.5}$" lines for ``CoSaMP" in Figure~\ref{fig:PlotW} appear beyond the scope of vertical axis. It can be seen that our metric-optimal algorithm outperforms the other two methods.

\begin{figure}[t]
\centering
	\begin{subfigure}[t]{0.3285\textwidth}
		\centering
		\includegraphics[width=55mm]{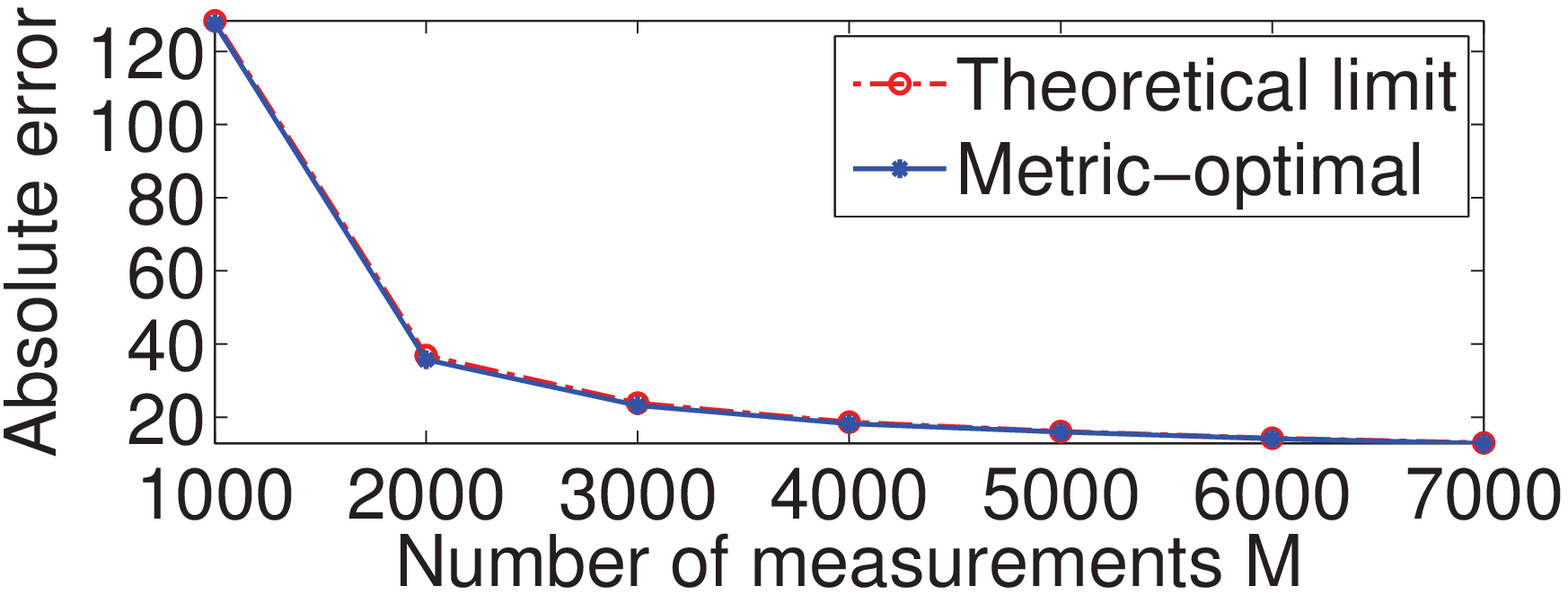}
\caption{\footnotesize\sl Absolute error.}
		\label{fig:plot_ell1}
	\end{subfigure}
	\begin{subfigure}[t]{0.3285\textwidth}
		\centering
		\includegraphics[width=55mm]{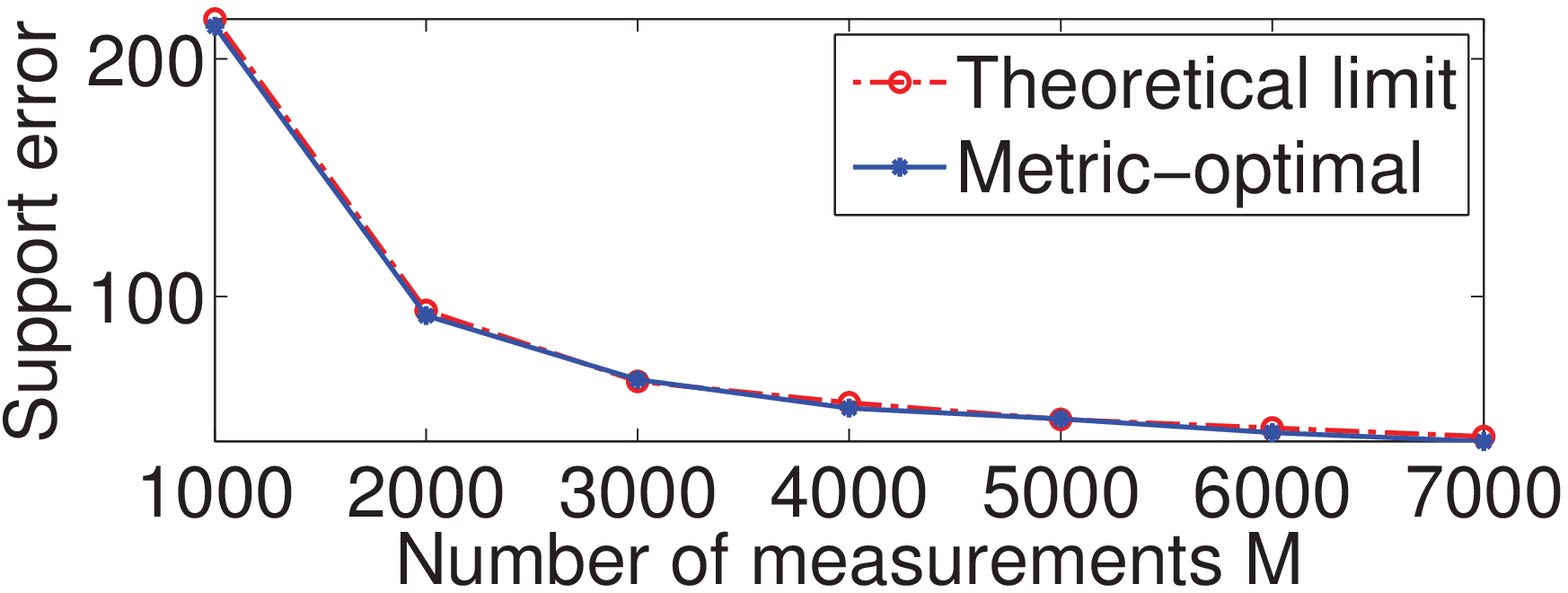}
\caption{\footnotesize\sl Support error.}
		\label{fig:plot_ell0}
	\end{subfigure}
	\begin{subfigure}[t]{0.3285\textwidth}
		\centering
		\includegraphics[width=55mm]{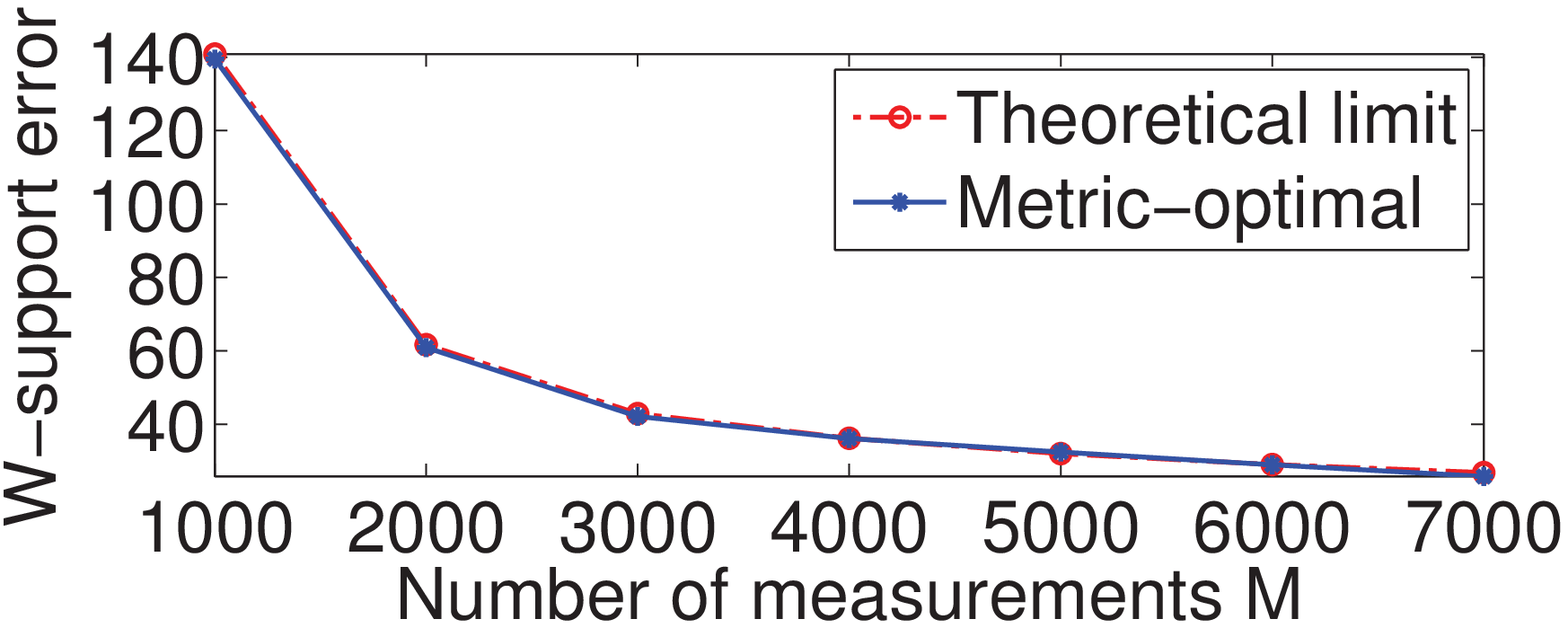}
\caption{\footnotesize\sl Weighted-support error.}
		\label{fig:plot_Well0}
	\end{subfigure}
\caption{
{\small\sl
Comparisons of the metric-optimal estimators and the corresponding theoretical limits~\eqref{eq:MMAEBound},~\eqref{eq:supportBound}, and~\eqref{eq:w_supportBound}. The corresponding two lines are on top of each other. (Sparse Gaussian input and Gaussian channel; sparsity rate $=3\%$; input length $N=10,000$; SNR $=20$ dB.)
}
\label{fig:Plot_theo}
}
\vspace*{-5mm}
\end{figure}

To demonstrate the theoretical analysis of our algorithm in Sections~\ref{subsec:Eg}, we compare our MMAE estimation results with the theoretical limit~\eqref{eq:MMAEBound} in Figure~\ref{fig:plot_ell1}, where the integrations are computed numerically. In Figure~\ref{fig:plot_ell0}, we compare our MMSuE estimator with the theoretical limit~\eqref{eq:supportBound}, where the value of~$\mu$ is acquired numerically from the relaxed BP method~\cite{Rangan:web:GAMP} with 20 iterations. 
In Figure~\ref{fig:plot_Well0}, our MMWSE estimator and its theoretical limit~\eqref{eq:w_supportBound} are compared, where we fix the weight~$\beta=0.3$, and obtain the value of~$\mu$ as in Figure~\ref{fig:plot_ell0}.
In all three figures, each point on the ``metric-optimal" line  is generated by averaging 40 experiments with the same parameters. It is shown from all figures that the two lines are on top of each other. Therefore our estimation algorithm reaches the corresponding theoretical limits and is optimal.

\begin{figure}[t]
\centering
\includegraphics[width=70mm]{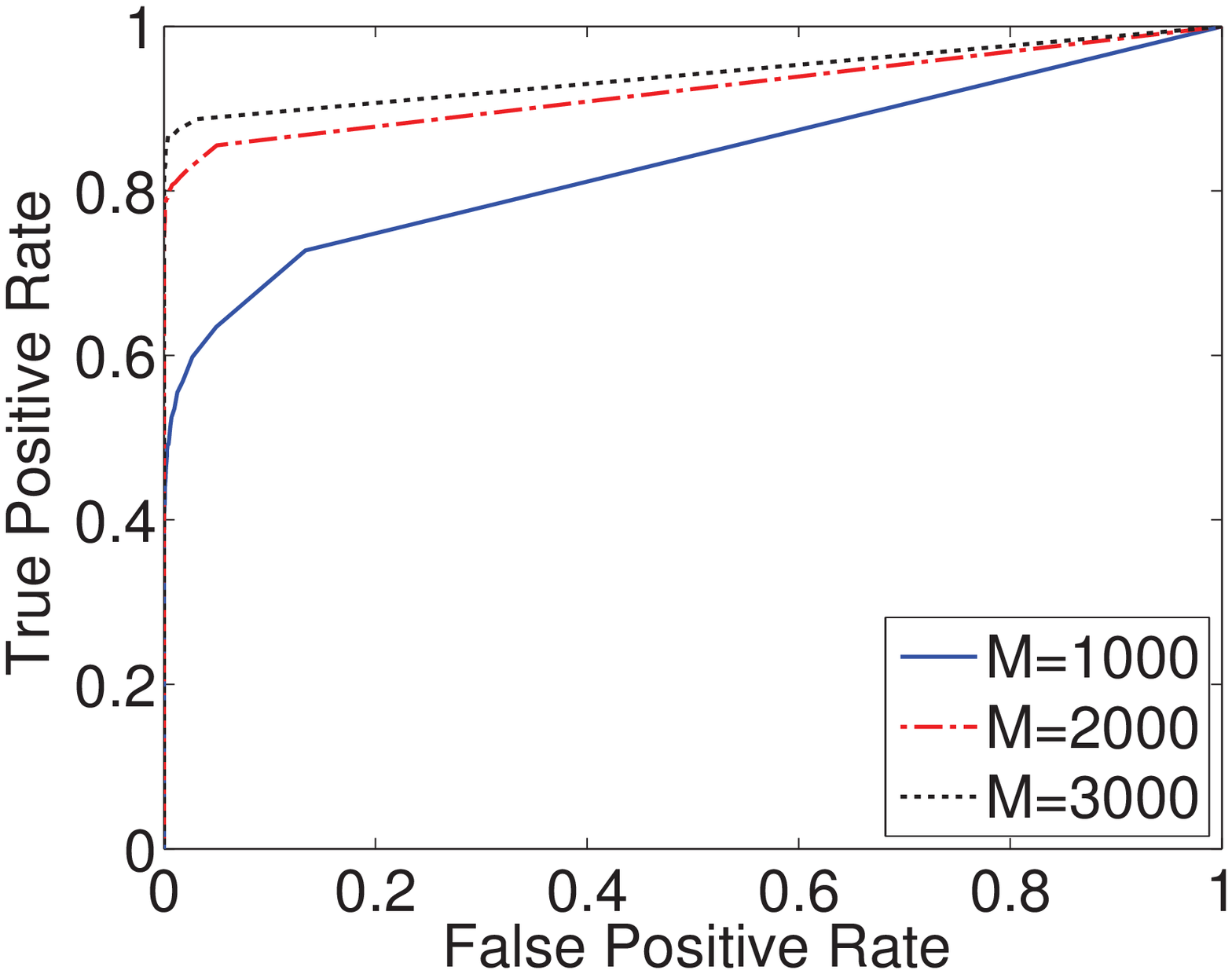}
\vspace*{-5mm}
\caption{
{\small\sl
The ROC curve obtained by setting weighted-support error as the error metric. (Sparse Gaussian input and Gaussian channel; sparsity rate $=3\%$; input length $N=10,000$; SNR $=20$ dB.)
}
\label{fig:plot_ROC}
}
\vspace*{-5mm}
\end{figure}

Figure~\ref{fig:plot_ROC} illustrates the ROC curve obtained by setting the weighted-support error~\eqref{eq:W_support} as the error metric. We vary the value of $\beta$ in~\eqref{eq:W_support} from $0$ to $1$, and compute the false positive rate as well as the false negative rate from~\eqref{eq:FAR} and~\eqref{eq:MDR}. The ROC curve is a $\Pr(\widehat{x}_j\neq0|x_j=0)$~\eqref{eq:FAR} versus $\Pr(\widehat{x}_j\neq0|x_j\neq0)$ plot, where $\Pr(\widehat{x}_j\neq0|x_j\neq0)$ is called the {\em true positive rate}, and~$\Pr(\widehat{x}_j\neq0|x_j\neq0)=1-\Pr(\widehat{x}_j=0|x_j\neq0)$~\eqref{eq:MDR}. In order to obtain different curves, we tune the number of measurements $M$, while keeping the sparsity rate, input length, and the SNR fixed. It can be seen that for the same level of false positive rate, a greater number of measurements achieves a higher true positive rate.

\section{Conclusion}
\label{sec:Conclusion}
In this correspondence, we introduced a pointwise estimation algorithm that deals with arbitrary additive error metrics in noisy compressed sensing. We verified that the algorithm is optimal in the large system limit, and provided a general method to compute the minimum expected error achievable by any estimation algorithm for a user-defined 
additive
error metric. We started with the scalar Gaussian channel model of the relaxed BP algorithm and extended it to a method that is applicable to any user-defined 
additive
error metric. 
We discussed three error metric examples, absolute error, support error, and weighted-support error, and gave the  theoretical performance limits for them. We further obtained the ROC curve for the modeled system by minimizing the weighted-support error.
We also illustrated numerically that our algorithm reaches
the three example theoretical limits, and outperforms the relaxed  BP and the CoSaMP methods.

\appendices
\section{Proof of Corollary \ref{coro:MMAE}}
\label{appen:MMAEProof}
When
\begin{eqnarray}
d_{\text{AE}}(\widehat{x_j},x_j)=|\widehat{x_j}-x_j|,\nonumber
\end{eqnarray}
equations~\eqref{eq:mainAlg} or~\eqref{eq:scalarEst}
solve for the MMAE estimand, $\widehat{\mathbf{x}}_{\text{MMAE}}$. In order to find the minimum, we take the derivative of the expected function over $\widehat{x}_j$,
\begin{eqnarray}
\label{eq:Derivative02}
\left.\frac{d\left.E(|\widehat{x}_j-x_j|\right|q_j)}{d\widehat{x}_j}\right|_{\widehat{x}_j=\widehat{x}_{j,\text{MMAE}}}=0,
\end{eqnarray}
for each $j=\{1,2,\ldots,N\}$. But
\begin{eqnarray}
&&E(|\widehat{x_j}-x_j||q_j)=\int_0^{\infty} 
\text{Pr}(|\widehat{x_j}-x_j|>t|q_j)dt\nonumber\\
&=&\int_0^{\infty}\text{Pr}(\widehat{x}_j-x_j>t|q_j)dt
+\int_0^{\infty}\text{Pr}(\widehat{x}_j-x_j<-t|q_j)dt\nonumber\\
&=&\int_{-\infty}^{\widehat{x}_j}\text{Pr}(x_j<t_1|q_j)dt_1+
\int_{\widehat{x}_j}^{\infty}\text{Pr}(x_j>t_2|q_j)dt_2,
\label{eq:MMAEMedian}
\end{eqnarray}
where changes of variables $t_1=\widehat{x}_j-t$ and $t_2=\widehat{x}_j+t$ are applied in \eqref{eq:MMAEMedian}. Using \eqref{eq:Derivative02} and \eqref{eq:MMAEMedian}, we need
\begin{eqnarray}
\text{Pr}(x_j<\widehat{x}_{j,\text{MMAE}}|q_j)-\text{Pr}(x_j>\widehat{x}_{j,\text{MMAE}}|q_j)=0,\nonumber
\end{eqnarray}
and thus $\widehat{x}_{j,\text{MMAE}}$ is given as the median of the conditional statistics $f_{X_j|Q_j}(x_j|q_j)$,
\begin{eqnarray}
\int_{-\infty}^{\widehat{x}_{j,\text{MMAE}}}f_{X_j|Q_j}(x_j|q_j)dx_j=\int_{\widehat{x}_{j,\text{MMAE}}}^{+\infty}f_{X_j|Q_j}(x_j|q_j)dx_j=\frac{1}{2}.\nonumber
\end{eqnarray}
Then, the conditional mean absolute error is,
\begin{eqnarray}
&&E[\left.|\widehat{x}_{j,\text{MMAE}}-x_j|\right|q_j]\nonumber\\
&=&\int_{-\infty}^{+\infty} |\widehat{x}_{j,\text{MMAE}}-x_j|f_{X_j|Q_j}(x_j|q_j)dx_j\nonumber\\
&=&\int_{-\infty}^{\widehat{x}_{j,\text{MMAE}}} (\widehat{x}_{j,\text{MMAE}}-x_j)f_{X_j|Q_j}(x_j|q_j)dx_j+\int_{\widehat{x}_{j,\text{MMAE}}}^{+\infty} (x_j-\widehat{x}_{j,\text{MMAE}})f_{X_j|Q_j}(x_j|q_j)dx_j\nonumber\\
&=&\int_{-\infty}^{\widehat{x}_{j,\text{MMAE}}} (-x_j)f_{X_j|Q_j}(x_j|q_j)dx_j+\int_{\widehat{x}_{j,\text{MMAE}}}^{+\infty} x_j f_{X_j|Q_j}(x_j|q_j)dx_j.\nonumber
\end{eqnarray}
Therefore, the MMAE for location $j$, $\text{MMAE}_j(f_{X_j},\mu)$, is
\begin{eqnarray}
&&\text{MMAE}_j(f_{X_j},\mu)=E[|\widehat{x}_{j,\text{MMAE}}-x_j|]\nonumber\\
&=&\int_{-\infty}^{+\infty} E[\left.|\widehat{x}_{j,\text{MMAE}}-x_j|\right|q_j]f_{Q_j}(q_j)dq_j\nonumber\\
&=&\int_{-\infty}^{+\infty}\left(\int_{-\infty}^{\widehat{x}_{j,\text{MMAE}}} (-x_j)f_{X_j|Q_j}(x_j|q_j)dx_j+\int_{\widehat{x}_{j,\text{MMAE}}}^{+\infty} x_j f_{X_j|Q_j}(x_j|q_j)dx_j\right)f_{Q_j}(q_j)dq_j.\nonumber
\end{eqnarray}
We note in passing that the integrations can be evaluated numerically in an implementation.

Because the input $\mathbf{x}$ is i.i.d., and the decoupled scalar channels have the same parameter $\mu$, the values of $\text{MMAE}_j$ for all $j\in\{1,2,\ldots,N\}$ are the same, and the overall MMAE is
\begin{eqnarray}
\text{MMAE}(f_{\mathbf{X}},\mu)=N\cdot\text{MMAE}_j(f_{X_j},\mu).\nonumber
\end{eqnarray}

\section{Proof of Corollary \ref{coro:MMSuE}}
\label{appen:MMSuEProof}
Similar to the idea of giving a limit on support recovery error rate \cite{Tulino2011}, we derive the MMSuE limit for the case where the input is real-valued and the matrix $\mathbf{\Phi}$ is rectangular ($M<N$). In the scalar Gaussian channel \eqref{eq:scalarGchannel}, we factor the sparse Gaussian input $X_j$ into $X_j=U_j\cdot B_j$, where $U_j\sim \mathcal{N}(0,\sigma^2)$ and $B_j\sim \text{Bernoulli}(p)$, i.e., $\text{Pr}(B_j=1)=p=1-\text{Pr}(B_j=0)$. The support recovery problem is the task of finding the maximum a-posteriori (MAP) estimation of $B_j$.

For our estimation algorithm, the conditional expectation of support recovery error is,
\begin{eqnarray}
E\left[d(\widehat{x_j},x_j)|q_j\right]=
\begin{cases}
\text{Pr}(B_j=1|q_j)&\text{if }\widehat{b_j}=0 \text{ and } b_j=1\\
\text{Pr}(B_j=0|q_j)&\text{if }\widehat{b_j}=1 \text{ and } b_j=0\\
0&\text{if }\widehat{b_j}=0 \text{ and } b_j=0\\
0&\text{if }\widehat{b_j}=1 \text{ and } b_j=1\nonumber
\end{cases}.
\end{eqnarray}
The estimand $\widehat{b}_{j,\text{opt}}$ minimizes $E\left[d(\widehat{x_j},x_j)|q_j\right]$, which implies
\begin{eqnarray}
\widehat{b}_{j,\text{opt}}=
\begin{cases}
0&\text{if }\text{Pr}(B_j=1|q_j)\le\text{Pr}(B_j=0|q_j)\\
1&\text{if }\text{Pr}(B_j=1|q_j)>\text{Pr}(B_j=0|q_j)
\end{cases}.
\label{eq:decisionRule01}
\end{eqnarray}
It is easy to see that $f_{Q_j|B_j}(q_j|0)\sim \mathcal{N}(0, \mu)$ and $f_{Q_j|B_j}(q_j|1)\sim \mathcal{N}(0, \sigma^2+\mu)$. Then,
\begin{eqnarray}
\text{Pr}(B_j=1|q_j)&=&\frac{f_{Q_j|B_j}(q_j|1)\text{Pr}(B_j=1)}{f_{Q_j}(q_j)}\nonumber\\
&=&\frac{f_{Q_j|B_j}(q_j|1)\text{Pr}(B_j=1)}{\sum_{b_j=0,1}f_{Q_j|B_j}(q_j|b_j)\text{Pr}(B_j=b_j)}\nonumber\\
&=&\frac{1}{1+\frac{1-p}{p}\sqrt{\sigma^2/\mu+1}\exp \left(-\frac{q_j^2}{2}\cdot\frac{\sigma^2/\mu}{\sigma^2+\mu}\right)},
\label{eq:Pr_b1}
\end{eqnarray}
and similarly
\begin{eqnarray}
\text{Pr}(B_j=0|q_j)&=&\frac{f_{Q_j|B_j}(q_j|0)\text{Pr}(B_j=0)}{f_{Q_j}(q_j)}\nonumber\\
&=&\frac{\frac{1-p}{p}\sqrt{\sigma^2/\mu+1}\exp \left(-\frac{q_j^2}{2}\cdot\frac{\sigma^2/\mu}{\sigma^2+\mu}\right)}{1+\frac{1-p}{p}\sqrt{\sigma^2/\mu+1}\exp \left(-\frac{q_j^2}{2}\cdot\frac{\sigma^2/\mu}{\sigma^2+\mu}\right)}.
\label{eq:Pr_b0}
\end{eqnarray}
Therefore, $\text{Pr}(B_j=1|q_j)>\text{Pr}(B_j=0|q_j)$ implies
\begin{eqnarray}
q_j^2>\tau=2\cdot\frac{\sigma^2+\mu}{\sigma^2/\mu}\ln\left(\frac{(1-p)\sqrt{\sigma^2/\mu+1}}{p}\right),\nonumber
\end{eqnarray}
and vice versa. We can rewrite \eqref{eq:decisionRule01} as,
\begin{eqnarray}
\widehat{b}_{j,\text{opt}}=
\begin{cases}
0&\text{if }q_j^2\le\tau\\
1&\text{if }q_j^2>\tau
\end{cases}.
\nonumber
\end{eqnarray}
By averaging over the range of $Q_j$, we get the overall MMSuE,
\begin{eqnarray}
&&\text{MMSuE}(f_{\mathbf{X}},\mu)=N\cdot E[d_{j,\text{support}}(\widehat{x_j},x_j)]\nonumber\\
&=&N\int E\left(d_{j,\text{support}}(\widehat{x_j},x_j)|q_j\right)f_{Q_j}(q_j)dq_j\nonumber\\
&=& N\int_{q_j^2>\tau}\text{Pr}(B_j=0|q_j)f_{Q_j}(q_j)dq_j+ N\int_{q_j^2\le\tau}\text{Pr}(B_j=1|q_j)f_{Q_j}(q_j)dq_j\nonumber\\
&=&N\cdot\text{Pr}(B_j=0,q_j^2>\tau)+N\cdot\text{Pr}(B_j=1,q_j^2\le\tau)\nonumber\\
&=&N\cdot\text{Pr}(q_j^2>\tau|B_j=0)\text{Pr}(B_j=0)+N\cdot\text{Pr}(q_j^2\le\tau|B_j=1)\text{Pr}(B_j=1)\nonumber\\
&=&N(1-p)\cdot\text{erfc}\left(\sqrt{\frac{\tau}{2\mu}}\right)+Np\cdot\text{erf}\left(\sqrt{\frac{\tau}{2(\sigma^2+\mu)}}\right).\nonumber
\end{eqnarray}

\section{Proof of Corollary \ref{coro:MMWSE}}
\label{appen:W_SuEProof}

We use the same variables $U_j$ and $B_j$ as defined in Appendix~\ref{appen:MMSuEProof}. For~$d_{\text{w\_support}}$~\eqref{eq:W_support}, its conditional expectation is
\begin{eqnarray}
E\left[d_{\text{w\_support}}(\widehat{x}_j,x_j)|q_j\right]=
\begin{cases}
(1-\beta)\cdot\text{Pr}(B_j=1|q_j)&\text{if }\widehat{b_j}=0 \text{ and } b_j=1\\
\beta\cdot\text{Pr}(B_j=0|q_j)&\text{if }\widehat{b_j}=1 \text{ and } b_j=0\\
0&\text{if }\widehat{b_j}=0 \text{ and } b_j=0\\
0&\text{if }\widehat{b_j}=1 \text{ and } b_j=1\nonumber
\end{cases}.
\end{eqnarray}
The estimand $\widehat{b}_{j,\text{opt}}$ minimizes $E\left[d_{\text{w\_support}}(\widehat{x_j},x_j)|q_j\right]$, which implies
\begin{eqnarray}
\widehat{b}_{j,\text{opt}}=
\begin{cases}
0&\text{if }(1-\beta)\cdot\text{Pr}(B_j=1|q_j)\le\beta\cdot\text{Pr}(B_j=0|q_j)\\
1&\text{if }(1-\beta)\cdot\text{Pr}(B_j=1|q_j)>\beta\cdot\text{Pr}(B_j=0|q_j)
\end{cases}.
\label{eq:bhat_Wsu}
\end{eqnarray}
Plugging~\eqref{eq:Pr_b1} and~\eqref{eq:Pr_b0} into~\eqref{eq:bhat_Wsu}, we get that
\begin{eqnarray}
\widehat{b}_{j,\text{opt}}=
\begin{cases}
0&\text{if }q_j^2\le\tau'\\
1&\text{if }q_j^2>\tau'
\end{cases},
\nonumber
\end{eqnarray}
where
\begin{eqnarray}
q_j^2>\tau'=2\cdot\frac{\sigma^2+\mu}{\sigma^2/\mu}\ln\left(\frac{\beta(1-p)\sqrt{\sigma^2/\mu+1}}{(1-\beta)p}\right).\nonumber
\end{eqnarray}
Therefore, the minimum mean weighted-support error function $\text{MMWSE}(f_{\mathbf{X}},\mu)$ is,
\begin{eqnarray}
&&\text{MMWSE}(f_{\mathbf{X}},\mu)=N\cdot E[d_{j,\text{w\_support}}(\widehat{x_j},x_j)]\nonumber\\
&=&N\int E\left(d_{j,\text{w\_support}}(\widehat{x_j},x_j)|q_j\right)f_{Q_j}(q_j)dq_j\nonumber\\
&=& N\int_{q_j^2>\tau'}\beta\text{Pr}(B_j=0|q_j)f_{Q_j}(q_j)dq_j+ N\int_{q_j^2\le\tau'}(1-\beta)\text{Pr}(B_j=1|q_j)f_{Q_j}(q_j)dq_j\nonumber\\
&=&N\beta(1-p)\cdot\text{erfc}\left(\sqrt{\frac{\tau'}{2\mu}}\right)+N(1-\beta)p\cdot\text{erf}\left(\sqrt{\frac{\tau'}{2(\sigma^2+\mu)}}\right).\nonumber
\end{eqnarray}
The false positive rate is
\begin{eqnarray}
\Pr(\widehat{x}_j\neq0|x_j=0)&=&\Pr(q_j^2>\tau'|x_j=0)\nonumber\\
&=&\Pr(q_j^2>\tau'|B_j=0)\nonumber\\
&=&\text{erfc}\left(\sqrt{\frac{\tau'}{2\mu}}\right),\nonumber
\end{eqnarray}
and the false negative rate is
\begin{eqnarray}
\Pr(\widehat{x}_j=0|x_j\neq0)&=&\Pr(q_j^2\le\tau'|x_j\neq0)\nonumber\\
&=&\Pr(q_j^2\le\tau'|B_j=1)\nonumber\\
&=&\text{erf}\left(\sqrt{\frac{\tau'}{2(\sigma^2+\mu)}}\right).\nonumber
\end{eqnarray}

\vspace*{0mm}
\section*{Acknowledgment}

We thank Sundeep Rangan for kindly providing the Matlab code \cite{Rangan:web:GAMP} of the relaxed BP algorithm \cite{Rangan2010,RanganGAMP2010}. We also thank the reviewers for their comments, which greatly helped us improve this manuscript.

\ifCLASSOPTIONcaptionsoff
  \newpage
\fi



\bibliography{cites}

\begin{thebibliography}{10}

\bibitem{Tan2012SSP}
J.~Tan, D.~Carmon, and D.~Baron,
\newblock ``Optimal estimation with arbitrary error metric in compressed
  sensing,''
\newblock in {\em Proc. IEEE Stat. Signal Process. Workshop (SSP)}, Aug. 2012,
  pp. 588--591.

\bibitem{TB2013ITA}
J.~Tan and D.~Baron,
\newblock ``Signal reconstruction in linear mixing systems with different error
  metrics,''
\newblock {\em Inf. Theory and App. Workshop (ITA)}, Feb. 2013.

\bibitem{DonohoCS}
D.~Donoho,
\newblock ``Compressed sensing,''
\newblock {\em IEEE Trans. Inf. Theory}, vol. 52, no. 4, pp. 1289--1306, Apr.
  2006.

\bibitem{CandesUES}
E.J. Candes and T.~Tao,
\newblock ``{Near-optimal signal recovery from random projections: Universal
  encoding strategies?},''
\newblock {\em IEEE Trans. Inf. Theory}, vol. 52, no. 12, pp. 5406--5425, Dec.
  2006.

\bibitem{CandesRUP}
E.~Cand\`{e}s, J.~Romberg, and T.~Tao,
\newblock ``Robust uncertainty principles: {E}xact signal reconstruction from
  highly incomplete frequency information,''
\newblock {\em IEEE Trans. Inf. Theory}, vol. 52, no. 2, pp. 489--509, Feb.
  2006.

\bibitem{BaraniukCS2007}
R.~G. Baraniuk,
\newblock ``A lecture on compressive sensing,''
\newblock {\em IEEE Signal Process. Mag.}, vol. 24, no. 4, pp. 118--121, July
  2007.

\bibitem{Rangan2010}
S.~Rangan,
\newblock ``Estimation with random linear mixing, belief propagation and
  compressed sensing,''
\newblock {\em CoRR}, vol. arXiv:1001.2228v1, Jan. 2010.

\bibitem{Grenander1957}
U.~Grenander and M.~Rosenblatt,
\newblock {\em Statistical analysis of stationary time series},
\newblock Wiley New York, 1957.

\bibitem{Levy2008}
B.C. Levy,
\newblock {\em Principles of signal detection and parameter estimation},
\newblock Springer Verlag, 2008.

\bibitem{Cover91}
T.~M. Cover and J.~A. Thomas,
\newblock {\em Elements of Information Theory},
\newblock Wiley-Interscience, 1991.

\bibitem{Webb2002}
A.R. Webb,
\newblock {\em Statistical pattern recognition},
\newblock John Wiley \& Sons Inc., 2002.

\bibitem{GuoVerdu2005}
D.~Guo and S.~Verd{\'u},
\newblock ``Randomly spread {CDMA}: {A}symptotics via statistical physics,''
\newblock {\em IEEE Trans. Inf. Theory}, vol. 51, no. 6, pp. 1983--2010, June
  2005.

\bibitem{Guo2006}
D.~Guo and C.C. Wang,
\newblock ``Asymptotic mean-square optimality of belief propagation for sparse
  linear systems,''
\newblock in {\em IEEE Inf. Theory Workshop}, Oct. 2006, pp. 194--198.

\bibitem{GuoBaronShamai2009}
D.~Guo, D.~Baron, and S.~Shamai,
\newblock ``A single-letter characterization of optimal noisy compressed
  sensing,''
\newblock in {\em Proc. 47th Allerton Conf. Commun., Control, and Comput.},
  Sep. 2009.

\bibitem{RFG2012}
S.~Rangan, A.~K. Fletcher, and V.~K. Goyal,
\newblock ``Asymptotic analysis of {MAP} estimation via the replica method and
  applications to compressed sensing,''
\newblock {\em IEEE Trans. Inf. Theory}, vol. 58, pp. 1902--1923, Mar. 2012.

\bibitem{CSBP2010}
D.~Baron, S.~Sarvotham, and R.~G. Baraniuk,
\newblock ``Bayesian compressive sensing via belief propagation,''
\newblock {\em IEEE Trans. Signal Process.}, vol. 58, pp. 269--280, Jan. 2010.

\bibitem{GuoWang2007}
D.~Guo and C.C. Wang,
\newblock ``Random sparse linear systems observed via arbitrary channels: A
  decoupling principle,''
\newblock in {\em Proc. Int. Symp. Inf. Theory (ISIT2007)}, June 2007, pp.
  946--950.

\bibitem{RanganGAMP2010}
S.~Rangan,
\newblock ``Generalized approximate message passing for estimation with random
  linear mixing,''
\newblock {\em Arxiv preprint arXiv:1010.5141}, Oct. 2010.

\bibitem{TroppOMP}
J.~A. Tropp and A.~C. Gilbert,
\newblock ``Signal recovery from random measurements via orthogonal matching
  pursuit,''
\newblock {\em IEEE Trans. Inf. Theory}, vol. 53, no. 12, pp. 4655--4666, Dec.
  2007.

\bibitem{Cosamp08}
D.~Needell and J.~A. Tropp,
\newblock ``Co{S}a{MP}: Iterative signal recovery from incomplete and
  inaccurate samples,''
\newblock {\em Appl. Comput. Harm. Anal.}, vol. 26, no. 3, pp. 301--321, May
  2009.

\bibitem{indyk2008near}
P.~Indyk and M.~Ruzic,
\newblock ``Near-optimal sparse recovery in the $\ell_1$ norm,''
\newblock in {\em 49th Annu. IEEE Symp. Found. Comput. Sci.}, Oct. 2008, pp.
  199--207.

\bibitem{berinde2008practical}
R.~Berinde, P.~Indyk, and M.~Ruzic,
\newblock ``Practical near-optimal sparse recovery in the $\ell_1$ norm,''
\newblock in {\em Proc. 46th Allerton Conf. Commun., Control, and Comput.},
  Sep. 2008, pp. 198--205.

\bibitem{CDDNOA}
A.~Cohen, W.~Dahmen, and R.~A. DeVore,
\newblock {\em Near optimal approximation of arbitrary vectors from highly
  incomplete measurements},
\newblock Inst. f{\"u}r Geometrie und Praktische Mathematik, 2007.

\bibitem{Wang2010}
W.~Wang, M.J. Wainwright, and K.~Ramchandran,
\newblock ``Information-theoretic limits on sparse signal recovery: Dense
  versus sparse measurement matrices,''
\newblock {\em IEEE Trans. Inf. Theory}, vol. 56, no. 6, pp. 2967--2979, June
  2010.

\bibitem{Tulino2011}
A.~Tulino, G.~Caire, S.~Shamai, and S.~Verd{\'u},
\newblock ``Support recovery with sparsely sampled free random matrices,''
\newblock in {\em IEEE Int. Symp. Inf. Theory (ISIT2011)}, July 2011, pp.
  2328--2332.

\bibitem{Wainwright2009}
M.J. Wainwright,
\newblock ``Information-theoretic limits on sparsity recovery in the
  high-dimensional and noisy setting,''
\newblock {\em IEEE Trans. Inf. Theory}, vol. 55, no. 12, pp. 5728--5741, Dec.
  2009.

\bibitem{Akcakaya2010}
M.~Ak{\c{c}}akaya and V.~Tarokh,
\newblock ``Shannon-theoretic limits on noisy compressive sampling,''
\newblock {\em IEEE Trans. Inf. Theory}, vol. 56, no. 1, pp. 492--504, Jan.
  2010.

\bibitem{Reeves2011sampling}
G.~Reeves and M.~Gastpar,
\newblock ``The sampling rate-distortion tradeoff for sparsity pattern recovery
  in compressed sensing,''
\newblock {\em IEEE Trans. Inf. Theory}, vol. 58, no. 5, pp. 3065--3092, May
  2012.

\bibitem{Bishop2006}
C.M. Bishop,
\newblock {\em Pattern recognition and machine learning},
\newblock Springer New York, 2006.

\bibitem{Caire2004}
G.~Caire, R.R. Muller, and T.~Tanaka,
\newblock ``Iterative multiuser joint decoding: Optimal power allocation and
  low-complexity implementation,''
\newblock {\em IEEE Trans. Inf. Theory}, vol. 50, no. 9, pp. 1950--1973, Sep.
  2004.

\bibitem{Montanari2006}
A.~Montanari and D.~Tse,
\newblock ``Analysis of belief propagation for non-linear problems: The example
  of {CDMA} (or: How to prove {T}anaka's formula),''
\newblock in {\em IEEE Inf. Theory Workshop}, Mar. 2006, pp. 160--164.

\bibitem{GuoWang2008}
D.~Guo and C.-C. Wang,
\newblock ``Multiuser detection of sparsely spread {CDMA},''
\newblock {\em IEEE J. Sel. Areas Commun.}, vol. 26, no. 3, pp. 421--431, Apr.
  2008.

\bibitem{Rangan2012}
S.~Rangan, A.K. Fletcher, V.K. Goyal, and P.~Schniter,
\newblock ``Hybrid generalized approximate message passing with applications to
  structured sparsity,''
\newblock {\em Proc. Int. Symp. Inf. Theory (ISIT2012)}, pp. 1236--1240, July
  2012.

\bibitem{Tanaka2002}
T.~Tanaka,
\newblock ``{A statistical-mechanics approach to large-system analysis of CDMA
  multiuser detectors},''
\newblock {\em IEEE Trans. Inf. Theory}, vol. 48, no. 11, pp. 2888--2910, Nov.
  2002.

\bibitem{DMM2009}
D.~L. Donoho, A.~Maleki, and A.~Montanari,
\newblock ``{Message passing algorithms for compressed sensing},''
\newblock {\em Proc. Nat. Acad. Sci.}, vol. 106, no. 45, pp. 18914--18919, Nov.
  2009.

\bibitem{Rangan:web:GAMP}
S.~Rangan, A.~Fletcher, V.~Goyal, U.~Kamilov, J.~Parker, and P.~Schniter,
\newblock ``{GAMP},''
  \url{http://gampmatlab.wikia.com/wiki/Generalized_Approximate_Message_Passing/}.

\end{thebibliography}

\end{document}